\documentclass{article}
\usepackage{spconf,amsmath,graphicx}

\usepackage{amsmath}
\usepackage{graphicx}
\usepackage[ruled,linesnumbered]{algorithm2e}
\usepackage{url}
\usepackage{multirow}
\usepackage{booktabs}
\usepackage{paralist}
\usepackage{color}
\usepackage{adforn}
\usepackage{colortbl}
\usepackage{xcolor}
\definecolor{darkblue}{rgb}{0, 0, 0.5}
\usepackage[colorlinks=true,
            linkcolor=red,
            anchorcolor=red,
            citecolor=blue,
            urlcolor=red
            ]{hyperref}

\definecolor{blue(pigment)}{rgb}{0.78, 0.4, 0.8}

\definecolor{Ocean}{RGB}{129,194,234}

\title{\textsc{After}: Active Learning Based Fine-Tuning Framework for Speech Emotion Recognition}
\name{BLIND}
\address{BLIND}
\name{Dongyuan Li, Yusong Wang, Kotaro Funakoshi, Manabu Okumura.}
\address{Tokyo Institute of Technology.}
\copyrightnotice{979-8-3503-0689-7/23/\$31.00~\copyright2023 IEEE}

\begin{document}

\maketitle

\begin{abstract}
Speech emotion recognition (SER) has drawn increasing attention for its applications in human-machine interaction.
However, existing SER methods ignore the information gap between the pre-training speech recognition task and the downstream SER task, leading to sub-optimal performance. 
Moreover, they require much time to fine-tune on each specific speech dataset, restricting their effectiveness in real-world scenes with large-scale noisy data.
To address these issues, we propose an active learning (AL) based Fine-Tuning framework for SER that leverages task adaptation pre-training (TAPT) and AL methods to enhance performance and efficiency.
Specifically, we first use TAPT to minimize the information gap between the pre-training and the downstream task. 
Then, AL methods are used to iteratively select a subset of the most informative and diverse samples for fine-tuning, reducing time consumption.
Experiments demonstrate that using only 20\%pt. samples improves 8.45\%pt. accuracy and reduces 79\%pt. time consumption.
\end{abstract}
\begin{keywords}
Speech Emotion Recognition, Large-scale Pre-trained Model, Fine-Tuning, Active Learning
\end{keywords}
\section{Introduction}
\label{sec:intro}

\textit{The language of tones is the oldest and most universal of all our means of communication}~\cite{blanton}. Speech emotion recognition (SER) aims to identify emotional states conveyed in vocal expressions as an essential topic in tone analysis. 
It has attracted much attention in both the industrial and academic communities, such as medical surveillance systems~\cite{clavel2008fear}, psychological treatments~\cite{elsayed2022speech,DBLP:conf/coling/LiYFO22}, and intelligent virtual voice assistants~\cite{la2020human}.

Emerging SER methods are broadly classified into Classic Machine Learning-based methods and Deep Learning-based methods~\cite{abbaschian2021deep}. The former methods~\cite{GharsellaouiSY19, Paraskevopoulos19,DBLP:conf/mir/WangLFO23} typically consist of three main components: feature extraction, feature selection, and emotion recognition. However, selecting and designing features for specific corpora is time-consuming~\cite{AyadiKK11}, and they always need better generalization on unseen datasets~\cite{PadiSSM21}. Deep learning-based methods can address these issues by automatically extracting more abstract features to improve generalization~\cite{LeCunBH15,DBLP:conf/icassp/MoraisHZGDA22,DBLP:journals/corr/abs-2307-10757}, benefiting from various neural network architectures such as convolutional neural networks (CNN)~\cite{Aftab} and Transformers~\cite{GhrissYRSW22}.
With the development of pre-trained language models~\cite{kenton2019bert} and the availability of large-scale datasets, various pre-trained automatic speech recognition (ASR) models, such as wav2vec 2.0~\cite{baevski2020wav2vec}, HuBERT~\cite{babu2021xls} and Data2vec~\cite{baevski2022data2vec}, have been proposed. These ASR models use speech's acoustic and linguistic properties to provide more robust and context-aware representations for speech signals. Xia et al.~\cite{XiaCRS21} proved that fine-tuning SER datasets on wav2vec 2.0~\cite{SchneiderBCA19} obtained state-of-the-art (SOTA) performance on IEMOCAP~\cite{BussoBLKMKCLN08}. This finding inspired researchers to explore new fine-tuning strategies on ASR models, which has become a new paradigm for SER. 
For example,
Ren et al.~\cite{abs-2210-14636} proposed a self-distillation SER model for fine-tuning wav2vec 2.0 and obtained SOTA performance on the DEMoS dataset~\cite{Cabaleiro20}.
Alef et al.~\cite{2207-14418} fine-tuned wav2vec 2.0 by jointly optimizing the SER and ASR tasks and achieving SOTA performance in Portuguese datasets.

Although the above methods have achieved great success, several issues still need to be solved. 1) current methods seldom consider the information gap between the pre-trained ASR and downstream SER task. For example, wav2vec 2.0~\cite{baevski2020wav2vec} adopts a masked learning objective to predict missing frames from the remaining context, while the downstream SER~\cite{Aftab, BaruahB22} task aims to minimize cross-entropy loss between predicted and referenced emotion labels for speech signals.
Suchin et al.~\cite{gururangan-etal-2020-dont} proved that the information gap would decrease the performance of downstream tasks.
To solve it, Pseudo-TAPT~\cite{2110-06309} first uses K-means to obtain pseudo-labels of speech signals and uses supervised TAPT~\cite{gururangan-etal-2020-dont} to continually pre-train.
However, K-means is sensitive to the initial value, making Pseudo-TAPT unstable and computationally expensive.
2) current methods only fine-tune and validate the performance on a specific speech dataset. For example, Xia et al.~\cite{XiaCRS21} used IEMOCAP, leading to over-fitting and poor generalization for unseen datasets. Real-world scenes contain much heterogeneous and noisy data, which hinders the application of these SER methods.
3) pre-trained ASR models often contain millions of parameters, such as wav2vec 2.0 contains 317 million parameters, which is time-consumption for real-world and large-scale datasets.

To alleviate the above issues, we propose an active learning-based fine-tuning framework for SER (\textsc{After}), which can be easily applied to noisy and heterogeneous real-world scenes.
Specifically,  we first propose an unsupervised task adaptation pre-training (TAPT) method~\cite{gururangan-etal-2020-dont} to reduce the information gap between the pre-trained and downstream SER task, where the pre-trained model can understand the semantic information of the SER task. 
Then, we created a large-scale heterogeneous and noisy dataset to simulate real-world scenes. Furthermore, we propose AL strategies with clustering-based initialization to iteratively select a smaller, more informative, and diverse subset of samples for fine-tuning, which can efficiently eliminate noise and outliers, improve generalization, and reduce the time-consuming.
Experimental results demonstrate the effectiveness and better generalization of \textsc{After} in noisy real-world scenes. Specifically, by fine-tuning only on 20\% pt. of the labeled samples, \textsc{After} can improve the unweighted accuracy by 8.45\%pt. compared to SOTA methods and reduce time consumption by 79\% pt. compared to the fastest baseline.




\begin{figure*}[ht]
\includegraphics[width=0.8\textwidth]{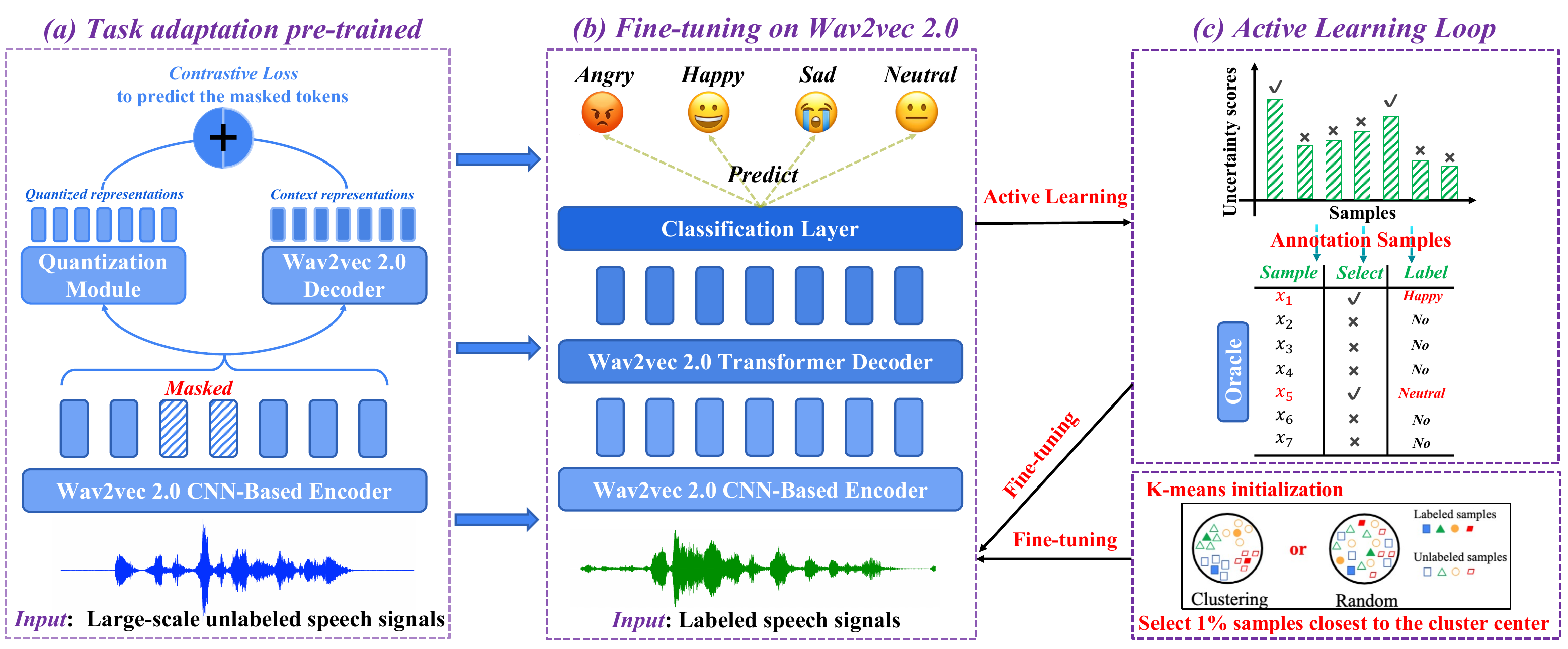}
\centering
\caption{Model overview. We first pre-train an off-the-shelf wav2vec 2.0 in the task adaptation pre-training manner. Then, we adopt an active learning method to select unlabeled samples for annotation iteratively. These labeled samples are used to fine-tune the wav2vec 2.0 model for speech emotion recognition.}
\label{overall}
\end{figure*}

\section{Methodology}
\label{method}
The overall framework is shown in Figure~\ref{overall}, where \textsc{After} contains three main components: a \textit{task adaptation pre-trained} module, an \textit{active learning-based fine-tuning} module and an \textit{emotion classification} module. 
First, we will formally give the task definition of SER and subsequently introduce each component of \textsc{After} in detail. 

\subsection{Task Formulation}
Given speech datasets $\mathcal{D}$ = $\{(\mathbf{x}_{i},y_{i})\}_{i=1}^{N}$ where $\mathbf{x}_{i}$ represents the $i$-th speech signal and $y_{i}$ represents its corresponding emotion label. 
We aim to fine-tune a pre-trained automatic speech recognition model $\mathbf{M}$, such as wav2vec 2.0~\cite{2110-06309}, on labeled speech datasets $\mathcal{D}$ to obtain accurately predicted emotion labels for all speech signals.

\subsection{Task Adaptation Pre-training (TAPT)}
Here we introduce our TAPT component in detail. 
To better leverage pre-trained prior knowledge for the benefit of downstream tasks, Gururangan et al.~\cite{gururangan-etal-2020-dont} continue training the pre-trained model RoBERTa~\cite{Roberta} on downstream datasets via the same loss of the pre-training task (reconstruct the masked tokens~\cite{kenton2019bert}) and significantly improved the performance on text classification.
Inspired by their work, we added an extra step to \textsc{After} by continuing the wav2vec 2.0 training speech recognition model on downstream SER's training datasets while keeping the same loss function with wav2vec 2.0 unchanged. By conducting this process, we bridge the information gap between the pre-trained ASR task and the target SER task, which we have preliminarily confirmed helps according to our experiments in Section~3.2.

As shown in Figure~\ref{overall} (a), the wav2vec 2.0 model $\mathbf{M}$($\mathbf{W}_{0}$), with pre-trained weights $\mathbf{W}_{0}$, consists of three sub-modules: the feature encoder module, the transformer module and the quantization module.
Specifically, we use a CNN-based encoder to encode the $i$-th input unlabeled speech signals into low-dimensional vectors as $\mathbf{x}_{i}$. Then, we randomly mask 15\%pt. features (following BERT~\cite{kenton2019bert}) of the speech vectors and decode them with two decoders to obtain quantized and context representations, where the quantization decoder can decode continuous speech vectors $\mathbf{x}_i$ into discrete code word from phonemes codebooks~\footnote{A quantized codebook refers to a set of predetermined values or codewords used to represent a continuous signal in a discrete form~\cite{baevski2020wav2vec}.} as $\mathbf{z}_{i}^{q}$ and wav2vec 2.0 decoder (transformer layers) can use self-attention to decode continuous speech vectors $\mathbf{x}_{i}$ into context-aware representations $\mathbf{z}_{i}^{c}$. 
Then, we design contrastive loss~\cite{baevski2020wav2vec} (cl) to minimize the differences between quantized and context representations as
\begin{equation}
\label{contrastive}
    \mathcal{L}_{cl} = -\sum_{i=1}^{n}\text{log}\frac{\text{exp}(\text{sim}(\mathbf{z}_{i}^{c}, \mathbf{z}_{i}^{q})/\kappa)}{\sum_{j=1}^{n}\text{exp}(\text{sim}(\mathbf{z}_{i}^{c}, \mathbf{z}_{j}^{q})/\kappa)},
\end{equation}
where the temperature hyperparameter $\kappa$ is set to 0.1,  and $\text{sim}(\mathbf{a}, \mathbf{b}) = \mathbf{a}^{T}\mathbf{b}/\|\mathbf{a}\|_{2}\|\mathbf{b}\|_{2}$ with $T$ representing the transposition of a vector.
Eq. (\ref{contrastive}) can help to obtain better quantized and context representations because two decoders can provide highly heterogeneous contexts for each speech signal~\cite{YouCSCWS20}.

To minimize the information gap between the pre-trained model and downstream SER task, following BERT~\cite{kenton2019bert}, we first randomly mask 15\%pt. tokens of each speech signal and then use reconstruction loss on the corrupted downstream SER dataset to generate tokens for reconstructing the original data, which can be formulated as
\begin{equation}
    \mathcal{L}_{rl} = - \frac{1}{|N_{m}|}\sum_{i=\text{First masked token}}^{\text{Last masked token}} s^{\text{true}}_{i}\text{log}(s_{i}^{\text{predicted}})
\end{equation}
where $N_{m}$ is the number of masked tokens, $s^{\text{true}}_{i}$ and $s_{i}^{\text{predicted}}$ are the ground-truth and predicted token probability of the $i$-th masked token.

Finally, we combine contrastive loss and reconstruction loss for the TAPT process as
\begin{equation}
    \mathcal{L}_{TAPT} = \mathcal{L}_{cl} + \mathcal{L}_{rl}.
\end{equation}

Although Pseudo-TAPT~\cite{2110-06309} also adopts TAPT, they spent much time using K-means to extract frame-level emotional pseudo labels and continually pre-train their model in a supervised manner. However, K-means is sensitive to the initial value and outliers~\cite{ZhangHJQAH20}, making Pseudo-TAPT unstable and computationally expensive.

\subsection{Active Learning (AL) based Fine-tuning}

After the TAPT process, we obtain the model $\mathbf{M}_{\textrm{TAPT}}(\mathbf{W}_{0}^{'})$ with $\mathbf{W}_{0}^{'}$ as the weight initialization for the AL process (cf. Line 1 of Algorithm~\ref{algorithm}). 
A typical AL setup starts by treating $\mathcal{D}$ as a pool of unlabeled data $\mathcal{D}_{\textrm{pool}}$ and performs $\tau$ iterations of sample selection. 
Specifically, in the $i$-th iteration, $k$ samples
$Q_{i}$ = $\{\mathbf{x}_{1},\cdots,\mathbf{x}_{k}\}$ are selected using a given acquisition function $ \mbox{ac}()$.
For example, we adopt Entropy~\cite{roy2001toward} as $\mbox{ac}()$ to measure the uncertainty of the samples and select the most uncertain $k$ samples. These selected samples are then labeled and added to the $i$-th training dataset $\mathcal{D}^{i}_{\textrm{train}}$, with which a model is fine-tuned for SER.

%
One primary goal of \textsc{After} is to explore whether AL strategies can reduce the number of annotation samples, as labeling large-scale datasets is the most laborious part of SER. Thus, for simplicity, we adopt five of the most well-known and influential AL strategies for evaluation,  including Entropy~\cite{roy2001toward}, Least Confidence~\cite{DredzeC08}, Margin Confidence~\cite{DredzeC08}, ALPs~\cite{YuanLB20} and BatchBald~\cite{KirschAG19}. These methods use different criteria to help select the most uncertain and informative samples from $\mathcal{D}_{\textrm{pool}}$. For example, we can adopt entropy to measure the uncertainty of $\mathbf{x}_{i}$ as \begin{equation} \textrm{Entropy}(\mathbf{x}_{i}) = -\sum_{j=1}^{c}P(y_{j}|\mathbf{x}_{i})\textrm{log}P(y_{j}|\mathbf{x}_{i}), \end{equation} where $c$ is the number of emotional classes and $P(y_{j}|\mathbf{x}_{i})$ represents the predicted probability of $\mathbf{x}_{i}$ for the $j$-th emotion.

And we select the number of $k$ most uncertainty samples for annotation and add them to the training dataset $\mathcal{D}_{\textrm{train}}$. Traditional AL methods use random initialization, while we found that the AL methods are sensitive to initialized samples and easily select redundant samples or outliers in each AL iteration with a bad initialization. Thus, instead of directly using AL methods, we propose a clustering-based initialization for all AL methods (we use K-means in this study) and obtain better performance (details about K-means are given in Sec~\ref{Active_learning_results}). Please note that, as shown in Algorithm~\ref{algorithm}, clustering-based initialization is only applied in the initialization process once, and the subsequent AL loop does not need a K-means process.

\begin{algorithm}
\small
\SetKwData{Left}{left}\SetKwData{This}{this}\SetKwData{Up}{up}
\SetKwFunction{Union}{Union}\SetKwFunction{FindCompress}{FindCompress}
\SetKwInOut{Input}{Input}\SetKwInOut{Output}{output}
\caption{Active Learning based Fine-tuning}
\label{algorithm}
\Input{Unlabeled data $\mathcal{D}_{\textrm{pool}}$, Model $\mathbf{M}(\mathbf{W}_{0})$, Acquisition size $k$, Iterations $\tau$, total number of selected samples $N_{c}$, and Acquisition function $\mbox{ac}()$.}
$\mathbf{M}_{\textrm{TAPT}}$ $(\mathcal{D}_{\textrm{pool}}; \mathbf{W}_{0}')$ $\leftarrow$ Train $\mathbf{M}(\mathbf{W}_{0})$ on $\mathcal{D}_{\textrm{pool}}$\;
$Q_{0} \leftarrow$ Clustering-based initialization from $\mathcal{D}_{\textrm{pool}}$\;
$\mathcal{D}_{\textrm{train}}^{0} \leftarrow Q_{0}$;
$\mathcal{D}_{\textrm{pool}}^{0} \leftarrow \mathcal{D}_{\textrm{pool}} \backslash Q_{0}$ where $|Q_{0}|=1\%N_{s}$\;
$\mathbf{M}_{0}([\mathbf{W}_{0}^{'},\mathbf{W}_{c}]) \leftarrow$ Initialized from $\mathbf{M}_{\textrm{TAPT}}$ $(\mathcal{D}_{\textrm{pool}}; \mathbf{W}_{0}^{'})$\;
$\mathbf{M}_{0}(\mathcal{D}_{\textrm{train}}^{0}; [\mathbf{W}_{0}^{'}, \mathbf{W}_{c}]) \leftarrow$ Train $\mathbf{M}_{0}([\mathbf{W}_{0}^{'}, \mathbf{W}_{c}])$ on $\mathcal{D}_{\textrm{train}}^{0}$\;
\For{i $\leftarrow$ 1 to $\tau$}
{$Q_{i}$ $\leftarrow$
$\mbox{ac}(\mathbf{M}_{i-1},\mathcal{D}_{\textrm{pool}}^{i-1},k)$ 
\quad \textcolor{blue}{$\triangleright$} \textcolor{blue}{Annotating $k$ samples}\;
$\mathcal{D}_{\textrm{train}}^{i}$\,\,=\,\,$\mathcal{D}_{\textrm{train}}^{i-1}$ $\cup$ $Q_{i}$  \,\, \textcolor{blue}{$\triangleright$} \textcolor{blue}{Add labeled samples to $\mathcal{D}_{\textrm{train}}^{i}$}\;
$\mathcal{D}_{\textrm{pool}}^{i} \leftarrow \mathcal{D}_{\textrm{pool}}^{i-1} \backslash Q_{i}$ \quad \textcolor{blue}{$\triangleright$} \textcolor{blue}{Delete samples from $\mathcal{D}_{\textrm{pool}}^{i}$}\;
$\mathbf{M}_{i}(\mathcal{D}_{\textrm{train}}^{i}; [\mathbf{W}_{0}^{'}, \mathbf{W}_{c}]) \leftarrow$ \textcolor{blue}{Train  $\mathbf{M}_{i-1}$ on $\mathcal{D}_{\textrm{train}}^{i}$}\; }
\end{algorithm}

\subsection{Emotion Recognition Classifier}

As shown in Figure.~\ref{overall} (b), we add a task-specific classification layer with additional parameters $\mathbf{W}_{c}$ for emotion recognition on the top of wav2vec 2.0. We fine-tune the classification model $\mathbf{M}_{i}([\mathbf{W}_{0}^{'}, \mathbf{W}_{c}])$ in each AL iteration with all labeled samples in $\mathcal{D}_{\textrm{train}}$ (cf. Lines 6-10 of Algorithm~\ref{algorithm}). We use the cross-entropy loss for the emotion recognition classifier:
\begin{equation}
    \mathcal{L}_{ce} = -\frac{1}{k} \sum_{i=1}^{k}\sum_{j=1}^{c} y^{j}_{i} \log{ (\hat{y}_{i}^{j})}, \label{cm3}
\end{equation}
where $c$ is the number of emotion classes,  $k$ is the number of slected samples at $t$-th iteration, $\hat{y}_{i}^{j}$ is the $i$-th predicted label, and $y_{i}^{j}$ is the $i$-th ground-truth of $j$-th class. 
 
\begin{table}[h]
\footnotesize
\caption{Descriptive statistics of the Merged dataset. Ratio of four labels is in the order of Anger : Neutral : Sad : Happy.}
\label{table1}
\centering
\setlength{\tabcolsep}{2mm}{
\begin{tabular}{@{}lcc@{}}
\toprule
\multirow{2}{*}{\textit{\textbf{Datasets}}} & \multicolumn{2}{c}{\textit{\textbf{Characteristics}}} \\ \cmidrule(l){2-3} 
& Number of Samples & Ratio of Four Labels  \\
\midrule 
\midrule
IEMOCAP~\cite{BussoBLKMKCLN08} (English) & 10,038  & \,2.5\,:\,1.2\,:\,2.4\,:\,1.0  \\
 EMODB~\cite{BurkhardtPRSW05} (German) & 408 & \,3.1\,:\,1.3\,:\,1.0\,:\,1.1   \\
 SHEMO~\cite{MohamadNezami2019} (English) & 2,737 & \,5.3\,:\,5.1\,:\,2.2\,:\,1.0  \\
 RAVDESS~\cite{0196391} (English) & 672 & \,2.0\,:\,1.0\,:\,2.0\,:\,2.0  \\ 
 EMov-DB~\cite{adigwe2018emotional} (English) & 3,038 & \,1.4\,:\,1.0\,:\,0.0\,:\,0.0  \\ 
 CREMA-D~\cite{CaoCKGNV14} (English) & 4,900 & \,1.0\,:\,1.7\,:\,1.0\,:\,1.0  \\ 
\midrule
\textbf{Merged Dataset} & 21,793 & \,1.5\,:\,1.4\,:\,1.0\,:\,1.5  \\
\bottomrule
\end{tabular}}
\end{table}

\section{Experiments and  Discussions}
\label{sec:pagestyle}

\subsection{Experimental Settings}
\label{subsection_experiment}

\subsubsection{Datasets}

We first evaluated the performance of all baselines using the widely used benchmark dataset, IEMOCAP~\cite{BussoBLKMKCLN08}. IEMOCAP is a multimodal database commonly employed for evaluating SER performance. There are five conversation sessions in IEMOCAP, each with a female and a male actor attempting to act in improvised and scripted scenarios. It consists of 10,039 speech utterances, with all audio signals sampled at 16kHz with a 16 bits resolution.
To ensure a fair comparison with previous works, we merge the ``excited'' class into the ``happy'' class, resulting in four considered emotions: neutral, happy, angry, and sad. Following Chen et al.~\cite{2110-06309}, we adopted a 5-fold cross-validation approach, where each IEMOCAP session was held out as the test set. We randomly selected 10\%pt. data from the remaining four sessions as our validation dataset and the rest as our training dataset.

Most existing methods are inadequate for real-world applications and susceptible to noise due to their heavy reliance on fine-tuning models using specific small-scale datasets. To address this issue, we conducted additional experiments by creating a larger training dataset. We achieved this by merging various datasets from different sources to simulate the noisy environments encountered in real-world scenarios. 
As shown in Table~\ref{table1}, we manually controlled the number of instances for each of the four labels in the Merged dataset to maintain label balance. 
Please note that EMODB is a German dataset that can help improve the noise of the merge dataset.
To explore whether the Merged dataset could improve performance on a single dataset, such as IEMOCAP, we also employed a 5-fold cross-validation approach by leaving each IEMOCAP session out as the test set, randomly selected 10\%pt. dataset from the remaining Merged dataset as our validation dataset, leaving the rest for training purposes.
Please note that we only employed the training data for both the TAPT and AL-based fine-tuning processes to prevent data leakage during evaluation. Furthermore, the training procedures were conducted from scratch separately for the IEMOCAP and Merged datasets.



\subsubsection{Baselines}

We compared various algorithms, including SOTA SER baselines and widely used AL methods. For SER methods, we selected the recently best-performing approaches: GLAM~\cite{9747517}, LSSED~\cite{9414542}, RH-emo~\cite{abs-2204-02385}, Light~\cite{Aftab}, Pseudo-TAPT~\cite{2110-06309}, and w2v2-L-r-12~\cite{10089511}. In terms of AL methods, we opted for the most efficient ones for our framework, which are: Entropy~\cite{roy2001toward}, Least Confidence~\cite{DredzeC08}, Margin Confidence~\cite{DredzeC08}, wav2vec 2.0 \&  clustering, ALPs~\cite{YuanLB20}, and BatchBald~\cite{KirschAG19}.

\subsubsection{Implementation details}

All experiments used the same learning rate $10^{-4}$ with the Adam optimizer. 
Our implementation of wav2vec 2.0 was based on the Hugging Face framework. The window length of the audio is set to 20 ms. 
We fine-tuned the model in a few-shot manner, which proposes longer fine-tuning, more evaluation steps during training, and early stopping with 20 epochs based on validation loss.
To have a fair comparison with the previous studies, we employed either off-the-shelf software packages or utilized the code provided by the respective authors. Each model was executed ten times, and the average performance across these runs was considered the final result. The choices of (hyper)parameters follow default if provided or tuned if not.
Following He et al.~\cite{10095777}, we evaluated the models using weighted accuracy (WA) and unweighted accuracy (UA)~\cite{metallinou2010decision} in speaker-independent settings. Please note that we did not require the data to be labeled by actual annotators. Instead,  we used the ground-truth labels available in the training dataset. Specifically, we masked the labels and only revealed them when the AL methods determined that the samples should be labeled, which is a common trick used by AL researchers to test out their ideas~\cite{roy2001toward}. However, it is worth mentioning that human annotators would be responsible for labelling the data in a real-world scenario.

\subsection{Active Learning Strategies Selection for \textsc{After}}
\label{Active_learning_results}

As shown in Figure~\ref{overall} (c), \textsc{After} incorporates an AL strategy for sample selection. To identify the most suitable AL method for \textsc{After}, we combined it with multiple well-known AL methods and evaluated their performance. Furthermore, we found that AL methods are sensitive to initialization, with most AL methods randomly selecting 1\%pt. samples for initialization~\cite{MargatinaVBA21}. Unlike them, we propose a novel clustering-based (K-means) initialization method to improve the performance of SER.
Specifically, we first extract sample representations of training data from the wav2vec 2.0 CNN-based encoder. Then, we employed K-means on the training data and selected 1\%pt. samples closest to the cluster centres as our initialized samples. Please note that we used elbow method~\cite{DBLP:journals/cin/SammoudaE21} to determine the number of clusters for K-means automatically, and we used the Euclidean distance to measure the distance between sample representations.

\begin{figure}[t]
\centering
\includegraphics[width=0.48\textwidth]{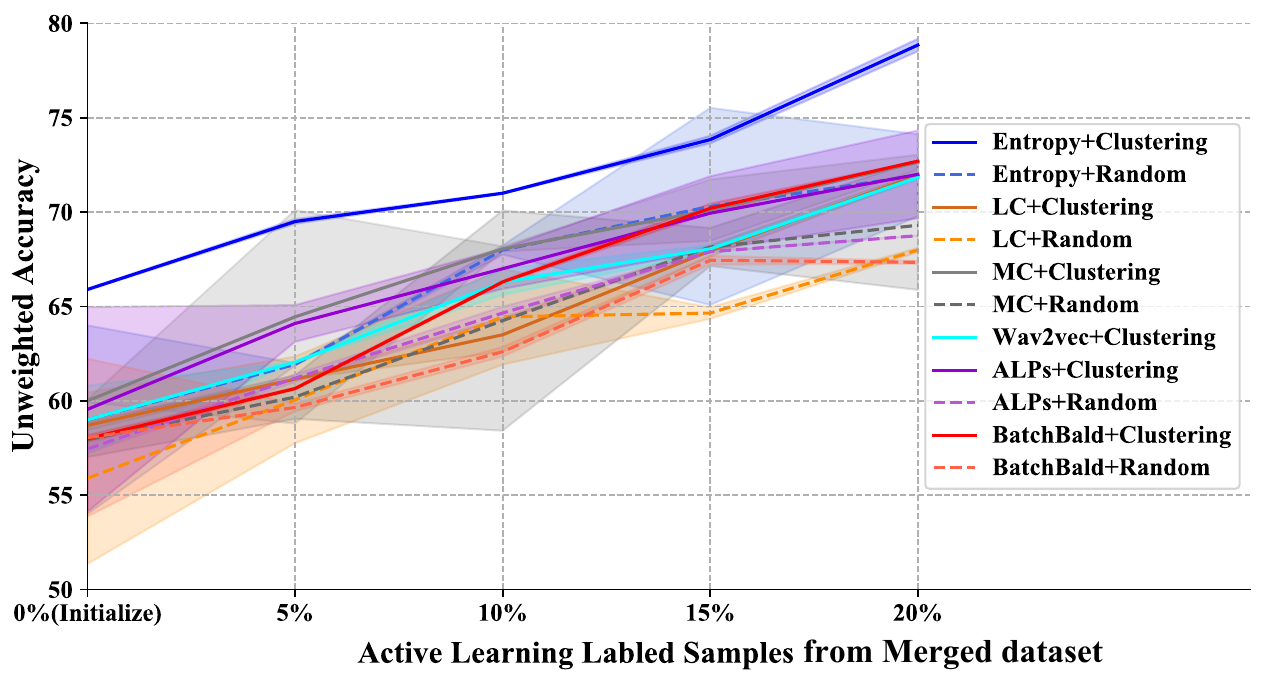}
\caption{Ratio of labeled samples vs. Unweighted Accuracy.}
\label{Ratio}
\end{figure}

Figure~\ref{Ratio} demonstrates that the clustering-based initialization outperforms the random initialization for all AL methods. 
The initial set of samples can well influence the selection order of samples in each iteration of AL, and an effective initialization can significantly enhance the performance and stability of AL methods. 
Figure~\ref{Ratio} illustrates that \textit{Entropy+Clustering} is the most effective AL strategy for \textsc{After} on the Merge dataset. 
Although we only show the diagram for UA due to space constraints, the diagram for WA is similar to UA.
\textbf{Therefore, \textit{Entropy+Clustering} is selected as the primary AL method for \textsc{After}}. 
And we recommend using \textit{Entropy+Clustering}, the simplest yet most efficient strategy for real-world applications.

\begin{table}[h]
\centering
\caption{\textsc{After} with Entropy~\cite{roy2001toward} to select 10\%pt.$\sim$100\%pt. labeled samples of the Merge Dataset for fine-tuning.}
\label{tab:tablenew}
\setlength{\tabcolsep}{1.2mm}{
\begin{tabular}{l*{6}{c}}
\toprule
\multicolumn{1}{l}{\multirow{2}{*}{\textbf{\textit{Metric}}}} 
& \multicolumn{6}{c}{\textbf{AFTER (TAPT+ AL-based FT)}}   \\ 
\cmidrule(r){2-7} 
 & 10\% & 20\%  & 40\%  & 60\%  & 80\%  & 100\%   \\
\midrule
\midrule
\textbf{UA}  & 71.45 & \textbf{77.41}  & 78.64 & \textbf{79.32} & 79.26 &79.15 \\
\textbf{WA}  & 69.01 & \textbf{74.32}  & 75.48 & \textbf{76.03} & 75.92 &75.94 \\
\textbf{Time} (mins) & 262.8 & \textbf{316.4}  & 785.4 & \textbf{942.2} & 1182.6 & 1508.2\\
\bottomrule
\end{tabular}
}
\end{table}

We analyzed the relationship between the ratio of labeled samples, performance, and time consumption of \textsc{After}. Results in Table~\ref{tab:tablenew} show that both performance and time consumption  of \textsc{After} increased as the ratio of labeled samples increased. \textbf{Our findings indicate that using 20\%pt. labeled samples yielded a significant improvement in performance while reducing the time consumption by \underline{79\%pt.} compared to fine-tuning on 100\%pt. samples.} Thus, we selected 20\%pt. labeled samples as a trade-off between performance and time consumption for subsequent experiments.

\begin{table}[h]
\centering
\caption{Overall performance comparison. 
\textsc{After} adopts Entropy+Clustering and selects 20\%pt. samples for fine-tuning. Symbol $\dag$ indicates \textsc{After} significantly surpass all baselines with $p<0.05$ according to t-test.}
\setlength{\tabcolsep}{1.9mm}{
\label{performance}
\begin{tabular}{@{}lcccc@{}}
\toprule
\multicolumn{1}{l}{\multirow{2}{*}{\textbf{\textit{Methods}}}} 
& \multicolumn{2}{c}{\textbf{\small{IEMOCAP}}} 
& \multicolumn{2}{c}{\textbf{\small{Merged Dataset}}} \\   
\cmidrule(lr){2-3} \cmidrule(lr){4-5} 
\multicolumn{1}{c}{}                         & UA $\uparrow$	         & WA  $\uparrow$       & UA  $\uparrow$        & WA $\uparrow$                   \\
\midrule
\midrule
GLAM [2022]                   & 74.01          & 72.98          & 71.38          & 69.28                          \\
LSSED [2021]                    & 73.09          & 68.35          & 25.00          & 36.20                          \\
RH-emo [2022]                    & 68.26          & 67.35          & 43.20          & 42.80                          \\
Light [2022]                    & 70.76          & 70.23          & 69.28          & 71.38                          \\
Pseudo-TAPT [2022]                    & 74.30          & 70.26          & 71.25          & 68.83       \\
w2v2-L-r-12 [2023] &74.28 &70.23 &71.22 &68.77\\
\midrule
\textbf{\textsc{After}}                                      & $ \,\,\textbf{76.07}^{\dag}$ & $ \,\,\textbf{73.24}^{\dag}$  & $ \,\,\textbf{77.41}^{\dag}$ &$ \,\,\textbf{74.32}^{\dag}$   \\
\bottomrule
\end{tabular}
}
\end{table}

\subsection{Comparison with Best-performing Baselines}

Table~\ref{performance} displays the primary results of \textsc{After} and the baselines on two datasets regarding unweighted and weighted accuracy. \textsc{After} outperforms all baselines with only 20\%pt. labeled samples for fine-tuning. Specifically, \textsc{After} improves UA and WA by \textbf{2.38\%pt. and 0.36\%pt.} respectively, compared to the SOTA baseline (UA of Pesdo-TAPT and WA of GLAM), on the Merged dataset. And \textsc{After} improves UA and WA by \textbf{\underline{8.45\%pt.} and 4.12\%pt.} respectively, compared to SOTA baseline (UA of GLAM and WA of Light), on the Merged dataset.

LSSED~\cite{9414542} and RH-emo~\cite{abs-2204-02385} achieved good results with the IEMOCAP but performed poorly with the Merged dataset. This discrepancy may be attributed to their limited denoising and domain transfer capabilities, preventing them from effectively handling the noise in the Merged dataset. On the other hand, GLAM~\cite{9747517} and Light~\cite{Aftab} employed multi-scale feature representations and deep convolution blocks to capture high-level global data features. They are beneficial for filtering out noisy low-level features and enhancing performance on both datasets. Psedp-TAPT~\cite{2110-06309} improved model robustness by using K-means to capture higher-level frame emotion labels as pseudo labels for supervised TAPT. Although baselines can eliminate the noise of the dataset to a certain extent, they exhibited high time complexity during fine-tuning with large-scale datasets. They did not effectively bridge the gap between pre-training and the downstream SER task. In contrast, \textsc{After} uses unsupervised TAPT to mitigate the information gap between the source domain (ASR) and the target (SER) domain. Additionally, \textsc{After} selects a subset of the most informative and diverse samples for iterative fine-tuning, which has three advantages: Firstly, it reduces labour consumption for manually labelling large-scale SER samples; Secondly, by utilizing a smaller labeled dataset, \textsc{After} significantly reduces the overall time consumption (Figure~\ref{time}), making it practical and feasible for real-world applications; Finally, the iterative fine-tuning process employed by \textsc{After} improves performance and stability by eliminating noise and outliers present in the selected samples, leading to enhanced overall model performance in SER tasks.

\begin{table}[h]
\footnotesize
\centering
\caption{Ablation study on the Merged dataset. FT means fine-tuning, and
TAPT+FT first adopted TAPT and then fine-tuned with  
corresponding selected labeled samples.
\textsc{After} adopts Entropy to select samples for fine-tuning.}
\setlength{\tabcolsep}{1.4mm}{
\label{ablation_study}
\begin{tabular}{@{}lcccccccc@{}}
\toprule
\multicolumn{1}{l}{\multirow{3}{*}{\textbf{\textit{Methods}}}} 
& \multicolumn{4}{c}{\textbf{\small{Random Sampling}}} 
& \multicolumn{4}{c}{\textbf{\small{Entropy Sampling}}} \\   
\cmidrule(lr){2-5} \cmidrule{6-9} 
\multicolumn{1}{c}{}                        & \multicolumn{2}{c}{\textbf{\small{FT}}}	         & \multicolumn{2}{c}{\textbf{\small{TAPT+FT}}}      &\multicolumn{2}{c}{\textbf{\small{FT}}}        & \multicolumn{2}{c}{\textbf{\small{\textsc{After}}}}                   \\
\cmidrule(lr){2-3} \cmidrule(lr){4-5}  \cmidrule(lr){6-7} \cmidrule{8-9}
\multicolumn{1}{c}{} & UA $\uparrow$ & WA $\uparrow$ & UA $\uparrow$ & WA $\uparrow$  & UA $\uparrow$ & WA $\uparrow$  & UA $\uparrow$ & WA $\uparrow$ \\
\midrule
\midrule
10\%pt. & 50.82 & 48.96 & 70.21 & 68.85 & 68.21 & 66.32 & 71.45 & 69.01 \\
20\%pt.          & 51.37 & 49.92 & 73.82 & 71.33
& 71.07   & 68.21   & \textbf{77.41}   & \textbf{74.32} \\
30\%pt.         & 52.37 & 50.18 &74.49 & 71.89      &72.35 & 69.21 & 78.20 & 75.16                \\
40\%pt.         & 55.68 & 52.21 & 76.01 & 72.28           &73.55 &70.18 & 78.64 & 75.48               \\
60\%pt.         & 60.39 & 59.32 & 77.21 & 74.58          &74.28 & 71.35 & \textbf{79.32} & \textbf{76.03}              \\
80\%pt.         & 58.34 & 56.72 & 78.88 & 75.82          &73.52 & 70.39 & 79.26 & 75.92               \\
100\%pt.                    & 57.21          & 54.12          & 78.21          & 75.36     & 73.89  & 70.89  & 79.15 & 75.94                    \\
\bottomrule
\end{tabular}
}
\end{table}

\subsection{Ablation Study for \textsc{After}}

We performed an additional ablation study to evaluate the efficacy of \textsc{After}, as shown in Table~\ref{ablation_study}. Specifically, we conducted fine-tuning (FT) and TAPT+FT on random sample selection and AL-based (Entropy) sample selection with varying ratios of labeled samples, ranging from 10\%pt. to 100\%pt..

From Table~\ref{ablation_study}, we have four interesting observations: (1) Fine-tuning with active learning significantly improves performance compared with random sampling (FT+Entropy vs FT+Random), regardless of the number of labeled samples. This result demonstrates that the AL-based fine-tuning strategy efficiently eliminates noise and outliers and selects the most informative and diverse samples for fine-tuning; (2) TAPT+FT outperforms FT on both random sampling and Entropy sampling, indicating that TAPT can effectively minimize the domain difference and significantly enhance the performance of the downstream SER task; (3) With the same number of labeled samples, \textsc{After} can obtain better results than TAPT+FT+Random on the Merged dataset. However, \textsc{After} with 20\%pt. labeled samples performs worse than TAPT+FT+Random with 80\%pt.$\sim$100\%pt. labeled samples. The reason is that TAPT+FT uses more labeled data for fine-tuning to prevent the model from overfitting and improve its robustness. In a fair comparison with the same size of the training data for fine-tuning, TAPT+FT+Random with 20\%pt. labeled samples performs worse than \textsc{After}(20\%pt.), demonstrating the effectiveness of \textsc{After}; (4) when 100\%pt. sample is used, AL-based method significantly outperforms random sampling method (FT+Random vs FT+Entropy). The main reason is that Random-sampling is affected by noise data, and the model constantly corrects the classification boundary, making it difficult to improve the results. Entropy sampling avoids the effect of noisy data by selecting the most informative and diverse samples for FT in advance to fix the classification boundary properly.

\subsection{Time Consumption Comparison}
\label{running_time}

\begin{figure}[t]
\centering
\includegraphics[width=0.5\textwidth]{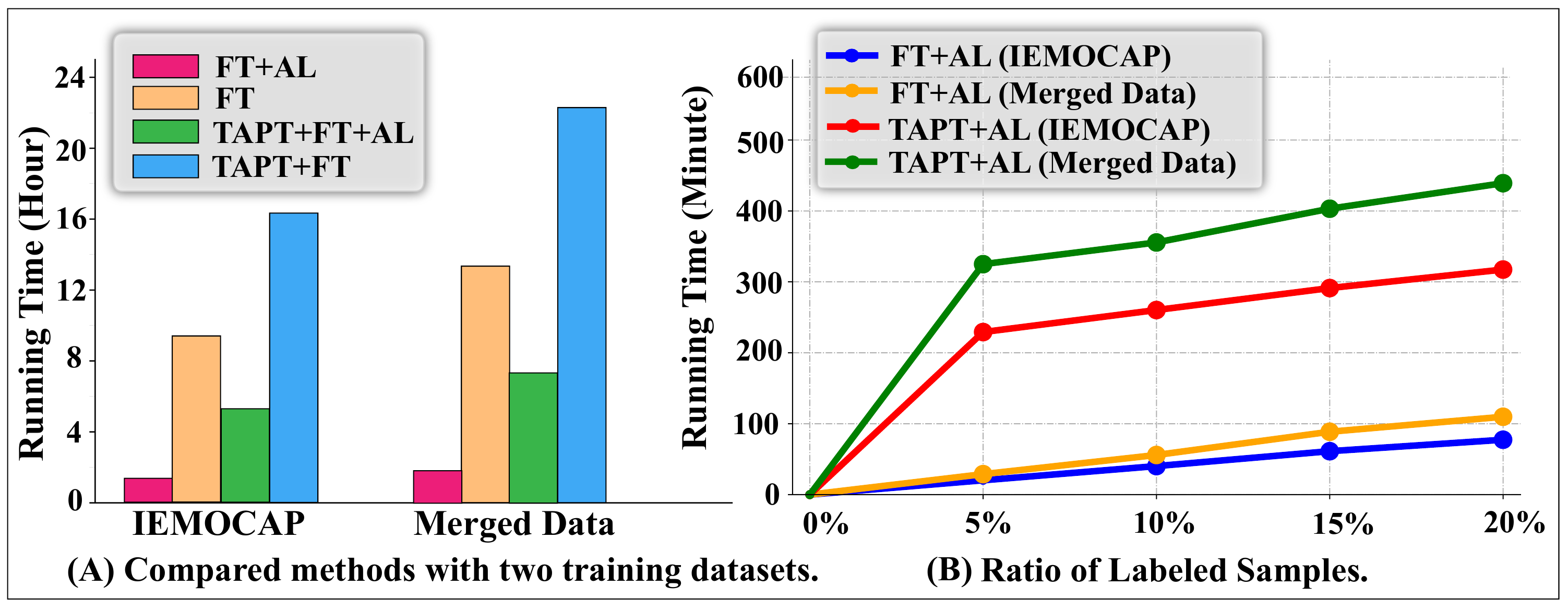}
\caption{ (A) Time Consumption Comparison and (B) The relationship between ratio of labeled samples and time consumption.}
\label{time}
\vspace{-0.7cm}
\end{figure}

Figure~\ref{time} (A) demonstrates that FT+AL with 20\%pt. labeled samples significantly reduces the time consumption of FT (fine-tuning on all labeled samples). Compared to TAPT+FT, TAPT+FT+AL significantly decreases the time consumption with the main cost incurred by TAPT. Additionally, the relationship between time consumption and the ratio of labeled samples is shown in Figure~\ref{time} (B). AL-based fine-tuning exhibits a linear increase in time consumption with sample size from 1\%pt.$\sim$20\%pt. (exponential growth from 30\%pt.$\sim$100\%pt. in Table~\ref{tab:tablenew}), indicating the efficiency of \textsc{After} and its potential to be applied in large-scale unlabeled real-world scenes.

\section{Conclusion and Future Work}
In this work, we investigated unsupervised TAPT and the AL-based fine-tuning strategy for improving the performance of SER. 
To extend SER for real-world applications, we constructed a large-scale noisy and heterogeneous dataset, and we used TAPT to minimize the information gap between the pre-trained and the target SER task. 
Experimental results indicated that \textsc{After} could dramatically improve performance and reduce time consumption.
We plan to design more domain adaptation AL methods to solve the SER task in the future.

\section*{Acknowledgement}
We would like to thank all the reviewers for their valuable suggestions, which helped us improve the quality of our manuscript.
And Dongyuan Li acknowledges the support from the China Scholarship Council (CSC).

\bibliographystyle{IEEEtran}
\bibliography{mybib}

\end{document}